# Graphene-protein bioelectronic devices with wavelength-dependent photoresponse

Ye Lu[1], Mitchell B. Lerner[1], Zhengqing John Qi[1], Joseph J. Mitala, Jr.[2§], Jong Hsien Lim[1,3], Bohdana M. Discher[2], A.T. Charlie Johnson[1*]

1 Department of Physics and Astronomy, University of Pennsylvania, Philadelphia PA   19104
2 Department of Biochemistry and Biophysics, University of Pennsylvania, Philadelphia PA   19104
3 Department of Department of Physics, Swarthmore College, Swarthmore, PA 19081

Correspondence to: A. T. Charlie Johnson email: cjohnson@physics.upenn.edu

**ABSTRACT** We implemented a nanoelectronic interface between graphene field effect transistors (FETs) and soluble proteins. This enables production of bioelectronic devices that combine functionalities of the biomolecular and inorganic components. The method serves to link polyhistidine-tagged proteins to graphene FETs using the tag itself. Atomic Force Microscopy and Raman spectroscopy provide structural understanding of the bio/nano hybrid; current-gate voltage measurements are used to elucidate the electronic properties. As an example application, we functionalize graphene FETs with fluorescent proteins to yield hybrids that respond to light at wavelengths defined by the optical absorption spectrum of the protein.

[§] Present address: Metabolism Branch, Center for Cancer Research, National Cancer Institute, National Institutes of health, Bethesda  MD  20892  USA

Graphene has drawn tremendous attention for its extraordinary electronic, mechanical and thermal properties [1]. Graphene also has potential for optical and optoelectronic applications, e.g, ultrafast photodetectors [2] and optical modulators [3]. For photo-detectors or -absorbers, it is desirable to control the wavelength of the device response, which is potentially problematic since pristine graphene monolayers show constant absorption of $\pi\alpha$=2.3%, where $\alpha$ is the fine structure constant, across the visible and infrared range [4]. This issue inspired our development of a reliable process to create hybrid bioelectronic devices that combine the functionality of biomolecules (here, proteins) with that of a graphene field effect transistor (GFET). When a fluorescent protein with an optical absorption peak at a particular wavelength is used, the GFET provides sensitive all-electronic readout of the protein's optical excitation. The approach thus enables the creation of a family of bio/nano hybrid photodetectors, each sensitive to a wavelength range defined by the proteinaceous component. The use of proteins with different functionalities (e.g., chemical affinity for particular biomarkers or small molecules in the liquid or vapor phase) could result in devices suitable for other applications, e.g., medical diagnostics or homeland security.

Experiments were performed on graphene produced by mechanical exfoliation onto oxidized silicon substrates. Graphene monolayers were selected by inspection with Atomic Force Microscopy (AFM) and Raman spectroscopy, as we have reported [5]. Devices were functionalized with carboxylated diazonium salts, which readily form covalent bonds with graphene [6]. As done for protein-carbon nanotube hybrids [7], the resulting carboxylic acid defects were activated with 1-ethyl-3-[3-dimethylaminopropyl]carbodiimide hydrochloride/sulfo- *N*-hydroxysuccinimide (EDC/sNHS), followed by attachment of nitrilotriacetic (NTA) [8]. Device fabrication was completed by adding Ni ions to the NTA complex, and incubation in protein solution (see Fig. 1(d)). Histidine-tagged proteins obtained commercially (protein-G (26.1 kDa), green fluorescent protein (GFP), and yellow fluorescent protein (YFP)) and a fusion protein were all attached to graphene devices, illustrating the robustness and versatility of the approach. The His-tagged (recombinant) fusion protein is comprised of glutathione S-transferase (GST), 6 histidine residues, and "BT5", an artificial heme binding protein [9].

Figure 1(a) and 1(b) are AFM images of the same graphene monolayer before and after functionalization with His-tagged protein-G, while Fig 1(c) shows height profiles along the indicated linescans. Two changes are observed: a ~2-nm increase in baseline height of the functionalized graphene, and the appearance of particles of height 3.4 ± 0.4 nm above the new baseline. In other experiments [10], AFM of graphene monolayers before and after incubation in diazonium solution showed height increases of ~ 0.5-nm height, in agreement with earlier reports[6], and we found a ~ 2-nm height increase upon attachment of NTA, with no particles observed (data not shown). Based on these observations, we associate the increased baseline height in Fig. 1(b) with a NTA layer and the particles with proteins bound by the His-tags. The claim of control over the point of the chemical bond to the protein in the hybrid and the device structure of Fig. 1(d) were confirmed by additional experiments: if the diazonium step is omitted, no molecules are found on the graphene surface, and if the protein lacks a His-tag, they do not bind to the Ni-NTA molecular layer.

For bio/nano hybrid devices, GFETs were fabricated using electron beam lithography. Care was taken to remove unwanted residues by thermal annealing as we reported previously [5,11]. Figure 2 shows current-gate voltage (I-$V_G$) data from a GFET before and after subsequent processing steps resulting in attachment of fusion protein GST-BT5. The annealed GFET I-$V_G$ shows ambipolar behavior and carrier mobility of ~2000 cm$^2$/V-s for holes and electrons. The neutrality point ($V_N$) occurs at gate voltage ~20V, corresponding to a doped carrier density of 1.6×10$^{12}$/cm$^2$ at $V_G$=0. After diazonium treatment (red dashed data), the device appears p-type with hole mobility ~300 cm$^2$/Vs; and $V_N$ exceeds 80V, the maximum gate voltage used. The carrier mobility decrease is attributed to defects formed by sp$^2$ bond breaking and attached carboxybenzene groups [12]; increased D-band intensity seen in the Raman spectrum supports this picture [13]. The $V_N$ shift is consistent with increased negative charge in the graphene environment, due to deprotonation of bound carboxybenzene groups in a nanoscale water layer formed under ambient. Subsequent attachment of Ni-NTA (blue dashed line) and GST-BT5 (green dotted line) does not affect the carrier mobility but the GFET conductance drops by ~25%, consistent with increased carrier scattering by bound molecules.

The protein-GFET bio/nano hybrid combines functionality of both components, similar to our reports on nucleic acid-GFET hybrids, where the nucleic acid provides chemical recognition and the GFET is used for sensitive electronic readout [11]. Here we explore integration of photoactive proteins to create hybrids with photoresponses at desired wavelength ranges.

Fig. 2(inset) shows I-$V_G$ data for GFP-GFET hybrid measured in the dark and when exposed to light of three different wavelengths (405 nm, 532 nm and 632 nm, referred to as violet, green, and red) at 70 mW/cm$^2$ intensity. A significant I-$V_G$ shift is observed only for violet illumination, with negligible change for green or red light. This wavelength-dependent photoresponse is consistent with the optical absorption spectrum of GFP, which is peaked near 400 nm with little absorption for wavelengths greater than 500 nm (inset to Fig. 3a; colored dots indicate wavelengths used). We attribute the I-$V_G$ shift to a GFET electronic response to the photoexcitation of GFP. The observed current decrease may reflect a net dipole associated with charge redistribution in GFP upon photoexcitation, or GFP-GFET charge transfer since GFP is reported to be a light induced electron donor [14]. We conducted control experiments to rule out the possibility that the photoresponse is due to the molecular layers alone [15].

Data in Fig. 3 provide further support for the production of bio/nano hybrids whose photocurrent response is determined by the optical properties of the protein component. Measurements were performed under ambient at $V_G = 0$ and $V_B = 10$ mV. In Fig. 3(a), starting at time 100 sec GFP-GFET hybrid was illuminated for 50 sec with light of a particular wavelength, and then the light was quenched for 50 sec; three cycles were used to gauge reproducibility. The response is shown as fractional change in DC current from the dark current baseline. For green and red illumination, the photoresponse is within the system noise (< 0.1%). In contrast, violet illumination induces a clear response of approximately -6%. We also showed that the wavelength of maximum hybrid device photoresponse is controlled by choice of fluorescent protein. For a yellow fluorescent protein (YFP)-GFET hybrid (Fig. 3(b)), strong conduction modulation occurs when the device is illuminated in the green, while excitation with violet or red light produces negligible response, as anticipated by the absorption spectrum of YFP (Fig. 3b, inset). Notably, FP-GFET devices are rather stable, with lifetimes exceeding two weeks [16].

To summarize, we developed a robust and reproducible method to bind His-Tagged proteins to graphene FETs. This creates a pathway for construction of bio/nano hybrids integrating desirable functionalities of both components. AFM, Raman spectroscopy, and control experiments were used to confirm the hybrid structure, and transport measurements to assess electronic effects of protein attachment. As an example of the capabilities enabled by the method, we demonstrated that FP-GFET hybrids form a class of photodetectors with photocurrent responses in a wavelength range determined by the absorption spectrum of the bound FP. Advances in design and synthetic control of proteins that fluoresce or incorporate other desirable functionality can be leveraged to develop quantitative understanding of mechanical, chemical, and electronic interactions in bio/nano hybrids, to monitor and stimulate protein biological activity, and to construct hybrid devices for applications in optoelectronics and chemical detection.

**Acknowledgement** We thank Tammer Farid (Dutton Group, Penn Biochemistry and Biophysics) for providing fusion protein GST-BT5; and Drs. Zhengtang Luo, Brett Goldsmith and Shawn Pfeil for useful discussions. Z.J.Q. acknowledges support from NSF IGERT # DGE-0221664. This work was supported by the Nano/Bio Interface Center (NBIC) through National Science Foundation NSEC DMR08-32802, and we acknowledge use of NBIC instrumentation.

Reference

1.  A.K.Geim and K.S.Novoselov, "The rise of graphene," Nature Materials **6**, 183-191 (2007).
2.  Thomas Mueller, Fengnian Xia, and Phaedon Avouris, "Graphene photodetectors for high-speed optical communications," Nat Photon **4** (5), 297-301 (2010); Fengnian Xia, Thomas Mueller, Yu-ming Lin, Alberto Valdes-Garcia, and Phaedon Avouris, "Ultrafast graphene photodetector," Nat Nano **4** (12), 839-843 (2009); Fengnian Xia, Thomas Mueller, Roksana Golizadeh-Mojarad, Marcus Freitag, Yu-ming Lin, James Tsang, Vasili Perebeinos, and Phaedon Avouris, "Photocurrent Imaging and Efficient Photon Detection in a Graphene Transistor," Nano Letters **9** (3), 1039-1044 (2009).
3.  Ming Liu, Xiaobo Yin, Erick Ulin-Avila, Baisong Geng, Thomas Zentgraf, Long Ju, Feng Wang, and Xiang Zhang, "A graphene-based broadband optical modulator," Nature **474** (7349), 64-67 (2011).
4.  R. R. Nair, P. Blake, A. N. Grigorenko, K. S. Novoselov, T. J. Booth, T. Stauber, N. M. R. Peres, and A. K. Geim, "Fine Structure Constant Defines Visual Transparency of Graphene," Science **320** (5881), 1308-1308 (2008).
5.  Y. Dan, Y. Lu, N.J. Kybert, Z. Luo, and A. T. C. Johnson, "Intrinsic response of graphene vapor sensors," Nano Lett. **9**, 1472-1475 (2009).
6.  Richa Sharma, Joon Hyun Baik, Chrisantha J. Perera, and Michael S. Strano, "Anomalously Large Reactivity of Single Graphene Layers and Edges toward Electron Transfer Chemistries," Nano Letters **10** (2), 398-405 (2010).
7.  B.R. Goldsmith, J.J. Mitala, J. Josue, A. Castro, M.B. Lerner, T.H. Bayburt, S.M. Khamis, R.A. Jones, J.G. Brand, S.G. Sligar et al., "Biomimetic chemical sensors using nanoelectronic readout of olfactory receptor proteins," ACS Nano **5**, 5408-5416 (2011); R.A. Graff, T.M. Swanson, and M. S. Strano, "Synthesis of Nickel-Nitrilotriacetic acid coupled single-walled carbon nanotubes for directed self-assembly with polyhistidine-tagged proteins," Chem. Mater. **20**, 1824-1829 (2008).
8.  See supplementary material at [URL will be inserted by AIP] for information on the suppliers of all chemical supplies and proteins used in the experiments.
9.  B.R. Lichtenstein, University of Pennsylvania, 2010.
10. See supplementary material at [URL will be inserted by AIP] for further discussion.
11. Y. Lu, B.R. Goldsmith, N.J. Kybert, and A. T. Charlie Johnson, "DNA decorated graphene chemical sensors," Appl. Phys. Lett. **97**, 083107 (2010).
12. Jian-Hao Chen, W. G. Cullen, C. Jang, M. S. Fuhrer, and E. D. Williams, "Defect Scattering in Graphene," Physical Review Letters **102** (23), 236805 (2009); Z. H. Ni, L. A. Ponomarenko, R. R. Nair, R. Yang, S. Anissimova, I. V. Grigorieva, F. Schedin, P. Blake, Z. X. Shen, E. H. Hill et al., "On Resonant Scatterers As a Factor Limiting Carrier Mobility in Graphene," Nano Letters **10** (10), 3868-3872 (2010).
13. See supplementary material at [URL will be inserted by AIP] for further discussion.
14. Alexey M. Bogdanov, Alexander S. Mishin, Ilia V. Yampolsky, Vsevolod V. Belousov, Dmitriy M. Chudakov, Fedor V. Subach, Vladislav V. Verkhusha, Sergey Lukyanov, and Konstantin A. Lukyanov, "Green fluorescent proteins are light-induced electron donors," Nat. Chem. Biol. **5** (7), 459-461 (2009).
15. See supplementary material at [URL will be inserted by AIP] for further discussion.
16. See supplementary material at [URL will be inserted by AIP] for further discussion.

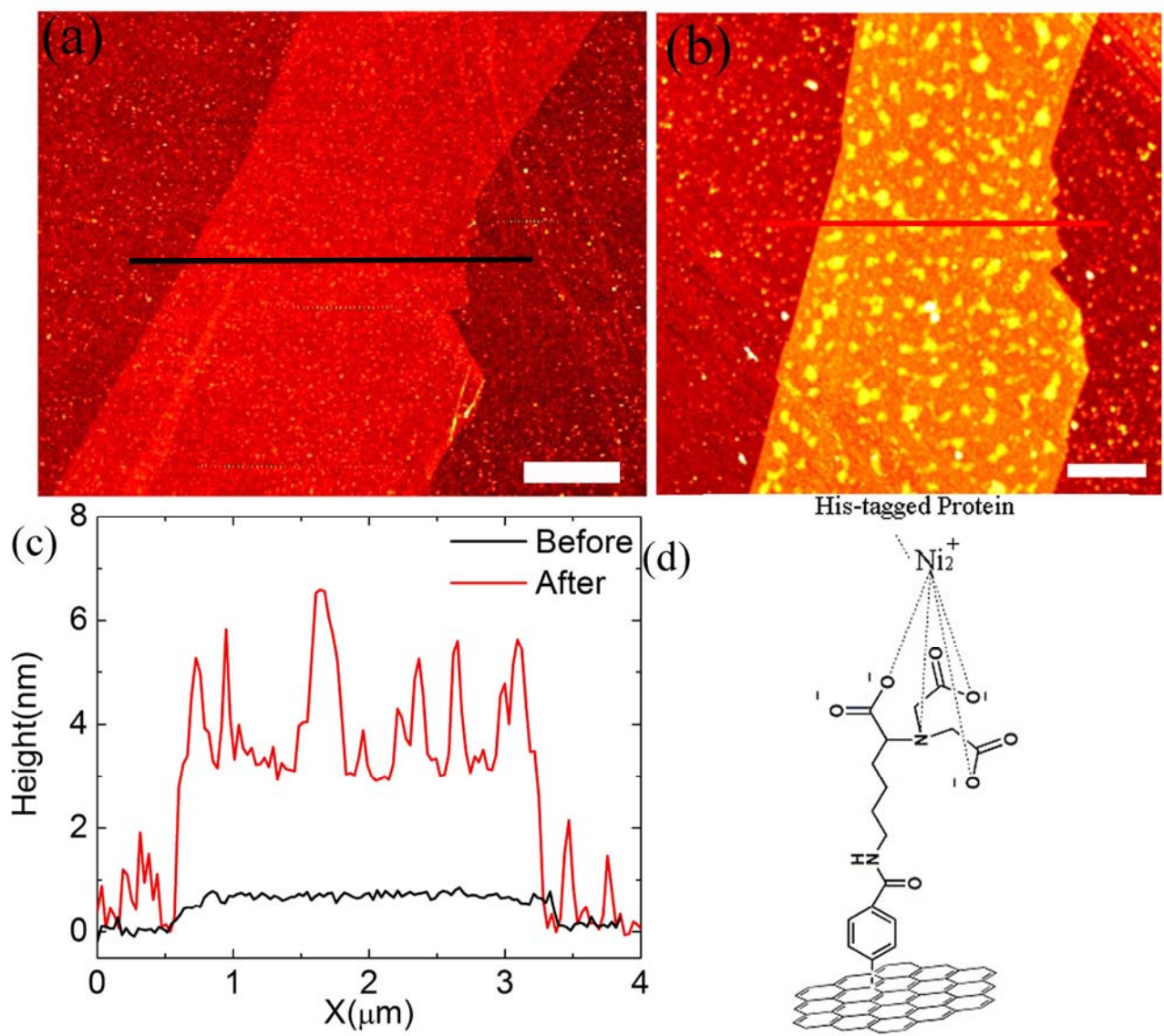

**Figure 1.** AFM of monolayer graphene before (a) and after (b) functionalization with His-tagged protein G (1μm scale bar for both images); z-scales are 8 nm and 20 nm, respectively), (c) Height linescans indicated on (a) and (b). (d) Schematic of chemical coupling between graphene (bottom) and the protein's histidine tag (top).

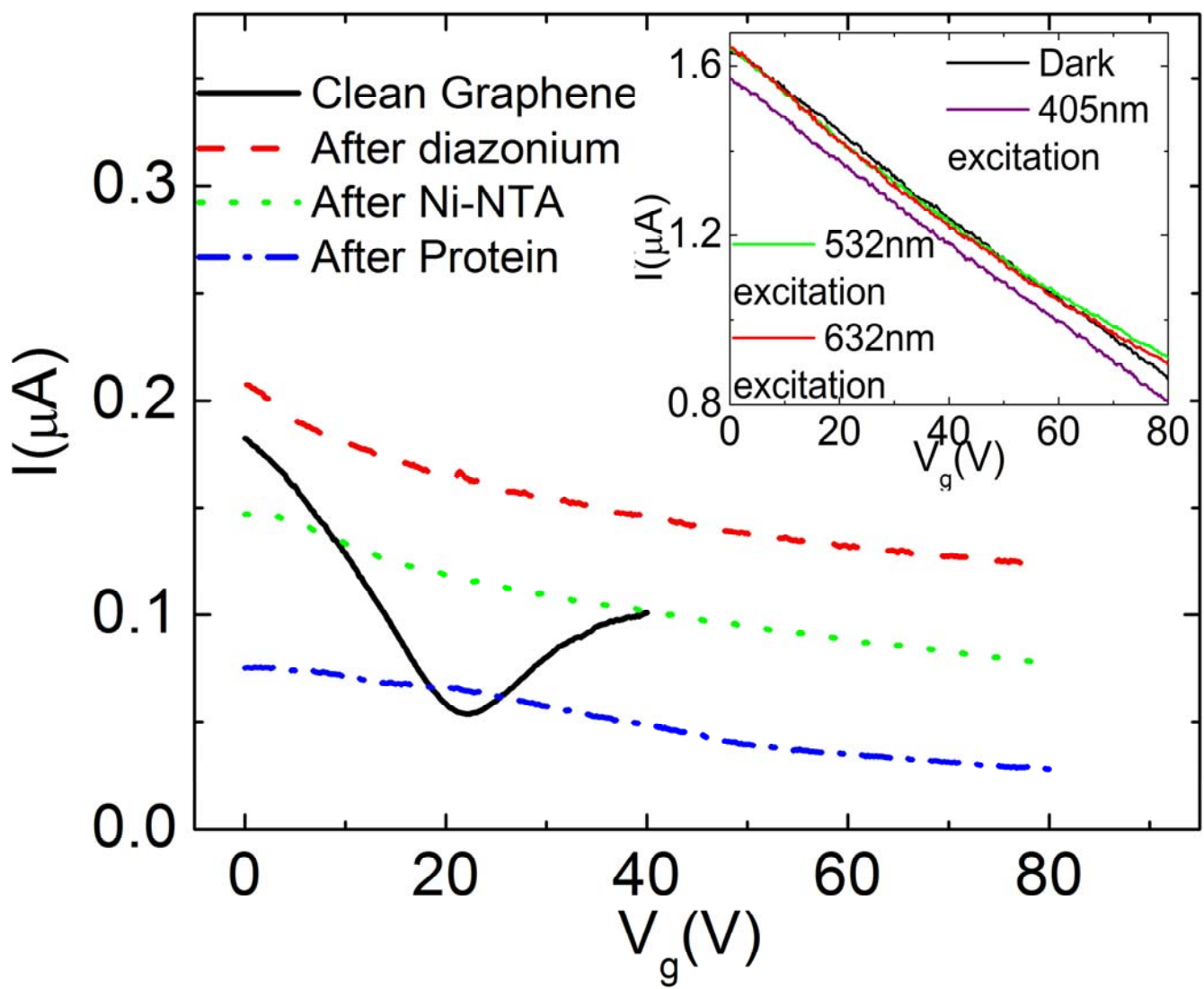

**Figure 2.** Current-gate voltage characteristic (I-$V_G$) of graphene FET after each step required for functionalization with fusion protein GST-BT5: as prepared (black), after diazonium treatment (red dashed), after Ni-NTA attachment (green dotted), and after incubation in protein (GST-BT5) solution (blue dot-dash). Bias voltage is 1 mV. (Inset) I-$V_G$ characteristics of a GFP-GFET with different illumination conditions: no illumination (black), and illumination at 405 nm (violet data), 532 nm (green data), and 632 nm (red data). Illumination intensity is approximately 70 mW/cm$^2$ for each wavelength. Bias voltage is 10 mV.

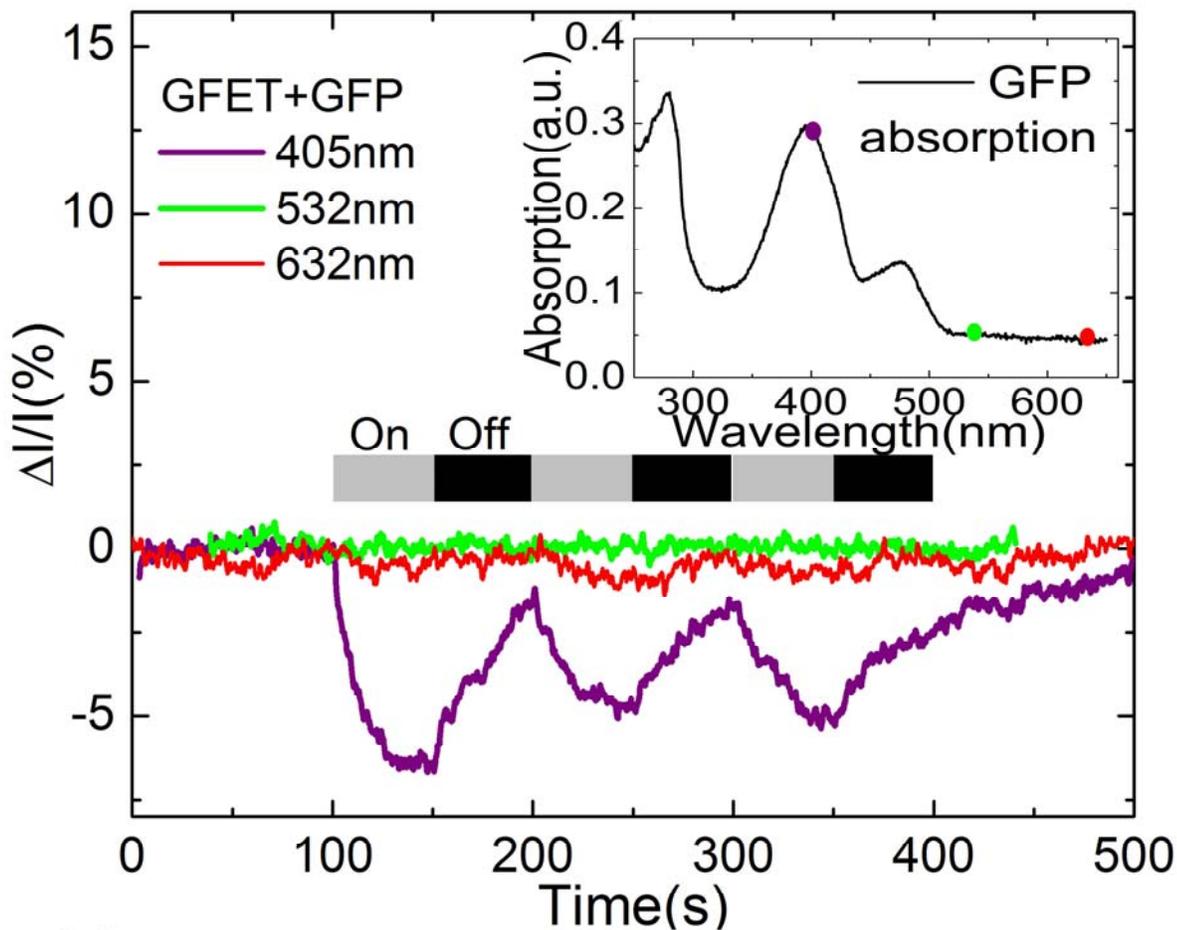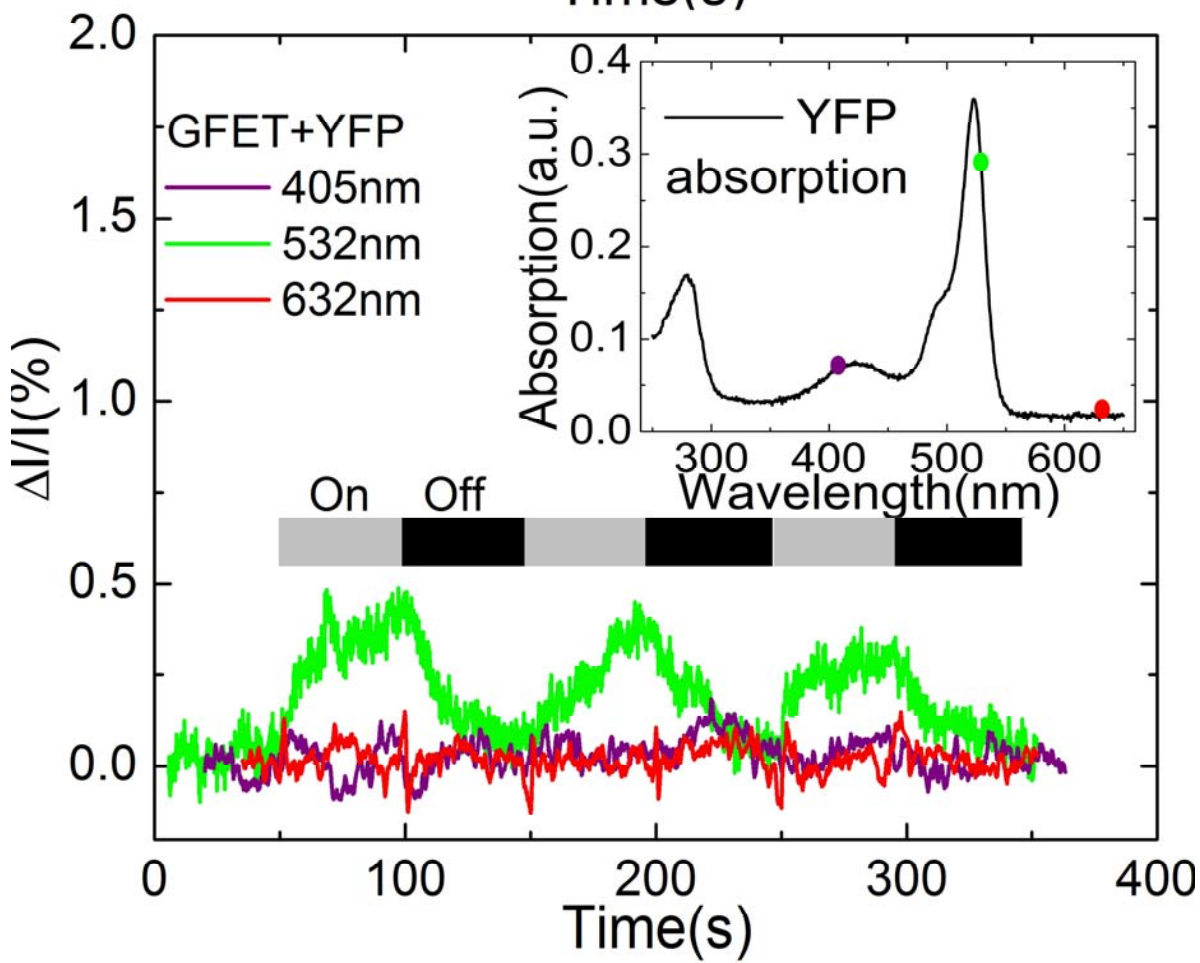

**Figure 3.** Photocurrent responses (%ΔI/I) of GFET hybrids incorporating green and yellow fluorescent protein (GFP and YFP, respectively) are determined by the proteins' optical absorption spectra. Responses are shown to illumination at 405 nm (violet data), 532 nm (green data), and 632 nm (red data). (a) Responses of GFP-GFET. Beginning at time 100 sec, the sample is illuminated for 50 sec and then the light is turned off for 50 sec. Only violet illumination causes a detectable response. Inset: Measured absorption spectrum of GFP. The wavelengths used in the experiments are indicated by appropriately colored dots. (b) Photocurrent responses of YFP-GFET hybrid. Beginning at time 50 sec, the sample is illuminated for 50 sec and then the light is turned off for 50 sec. Now only green illumination produces a detectable response. Inset: Measured absorption spectrum of YFP with dots indicating wavelengths used.